# Using survival curves for comparison of ordinal qualitative data in clinical studies

Basilio de Bragança Pereira[*,a,b], Emília Matos do Nascimento[a], Felippe Felix[b], Guilherme Ferreira da Motta Rezende[b]

[a]*Federal University of Rio de Janeiro, COPPE - Postgraduate School of Engineering, Rio de Janeiro, Brazil*

[b]*Federal University of Rio de Janeiro, School of Medicine and HUCFF - University Hospital Clementino Fraga Filho, Rio de Janeiro, Brazil*

---

**Abstract**

**Background and Objective:** The survival-agreement plot was proposed and improved to assess the reliability of a quantitative measure. We propose the use of survival analysis as an alternative non-parametric approach for comparison of ordinal qualitative data.

**Study Design and Setting:** Two case studies were presented. The first one is related to a randomized, double blind, placebo-controlled clinical trial to investigate the safety and efficacy of silymarin/metionin for chronic hepatitis C. The second one is a prospective study to identify gustatory alterations due to chorda tympani nerve involvement in patients with chronic otitis media without prior surgery.

**Results:** No significant difference was detected between the two treatments related to the chronic hepatitis C ($p > 0.5$). On the other hand, a significant association was observed between the healthy side and the affected side of the face of patients with chronic otitis media related to gustatory alterations ($p < 0.05$).

**Conclusion:** The proposed method can serve as an alternative procedure to statistical test for comparison of samples from ordinal qualitative variables. This approach has the advantage of being more familiar to clinical researchers.

[*] Corresponding author. Tel.: +55-21-25622594
E-mail address: basilio@hucff.ufrj.br (B.B. Pereira)

**What is New**

- A method for comparison of ordinal qualitative data using survival analysis is evaluated for the first time.

- A new application of a methodology more familiar to medical researchers.

- An alternative non-parametric method to test equality of two or more samples.

## 1. Introduction

In clinical research, recent studies have been evaluating in order to assess agreement of quantitative data. Luiz et al. [1] proposed the survival-agreement plot approach to assess the reliability of a quantitative measure. The authors expressed the degree of agreement (or disagreement) of a measure as a function of several limits of tolerance, using the Kaplan-Meier [2] method, where the failures occur exactly at absolute values of the differences between the two methods, say A and B. According to the authors, the survival-agreement plot is a step function of a typical survival analysis without censored data, where the Y axis represents the proportion of discordant cases. This is equivalent to a step function where the X axis represents the absolute observed differences and the Y axis is the proportion of the cases with at least the observed difference ($x_i$).

An improvement on the Luiz et al. [1] method was suggested by Llorca and Delgado-Rodríguez [3], by considering two groups of the real differences (A < B and A > B) instead of the absolute values of the observed difference between the methods (|A – B|). Therefore, usual tests for comparisons of survival curves can be evaluated.

In this paper we propose the use of survival analysis method for comparison of ordinal qualitative data.

## 2. Data base and methods

In this section two case studies are presented to illustrate the new approach by using survival analysis to assess agreement of the ordinal qualitative data.

The first one is related to chronic hepatitis C, which is a widespread infectious disease. The World Health Organization estimates that about 3% of the world's population has been infected with hepatitis C virus and that some 170 million are chronic carriers [4]. Although specific antiviral therapy with interferon alpha plus ribavirin is strongly recommended to patients with progressive disease, the overall response rate is less than fifty percent [5]. Non responders have high risk of developing progressive hepatic fibrosis, and might be beneficiated by reversing it through the use of anti-fibrotic drugs. One of these drugs is silymarin, obtained from milk thistle and used since ancient times as a liver tonic, alone or combined with metionin [6]. To evaluate the safety and efficacy of silymarin/metionin for chronic hepatitis C a randomized, double blind, placebo-controlled clinical trial was conducted, in which both groups were concurrently treated with interferon alpha plus ribavirin. The primary efficacy end point was improvement of liver histology after treatment. Changes in histology were evaluated comparing previous and pos therapy biopsy features, according to the Ishak grading system [7] that classifies the degree of inflammation in nineteen grades (0 to 18) and the extent of fibrosis in seven stages (0 to 6).

The second case study is related to 45 patients with chronic otitis media (COM), being 25 with cholesteatoma and 20 without cholesteatomatous COM, with a mean age of 38 years. Eight cases of unilateral ageusia were found on the affected side.

A prospective study was performed to identify gustatory alterations due to chorda tympani nerve involvement in patients with chronic otitis media (COM) without prior surgery, and to find out whether the presence of cholesteatoma worsened gustatory sensitivity in these patients. The test was performed in patients with unilateral cholesteatomatous or suppurated COM not previously submitted to otological surgery. The test was based on "taste strips" with different concentrations of salt, sweet, bitter, and sour, using the otological disease free side as the control. The score could be between 0 (worst) and 16 (best). The data were collected by interview and physical exam. Patients with cholesteatomatous or suppurated COM may present gustatory alterations, even in the absence of complaints.

## 3. Results

Figure 1 presents the survival curves related to 42 patients considered in the first case study, who were divided into two groups of treatments. These two groups were compared in the beginning and in the end of the study. The survivorship functions cross indicating that the proportional hazards assumption was violated. Therefore to test the equality of the two survivor functions the Tarone-Ware test [8] was performed, which may be more powerful non-parametric weighted rank test than the alternatives log rank test and the modified Wilcoxon test. The results indicated no significant difference between the groups before and after treatment ($p > 0.5$).

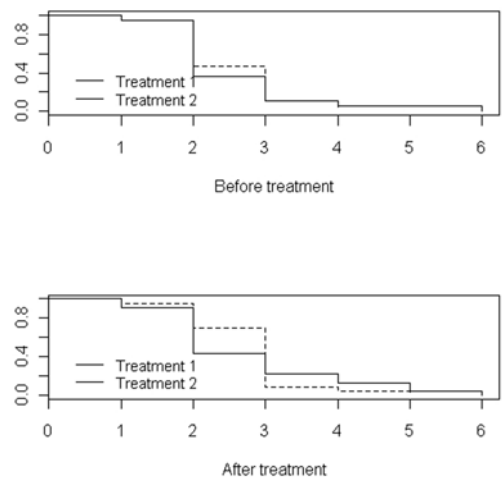

**Figure 1 – Survival curves before and after treatment**

In the second case study, related to patients with chronic otitis media (COM), a Poisson regression and the above survival curve analysis was used. The covariates considered in this study were: age, gender, smoking, otorrhea, cholesteatoma, diabetes, and blood pressure. A typical result of the survival curve is shown in Figure 2 for comparison of the healthy and affected side. After performing the Tarone-Ware test [8], the results pointed out significant association between the healthy and affected side ($p = 0.00239$).

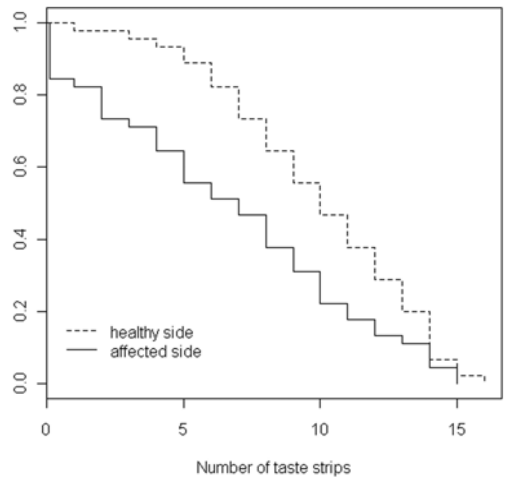

**Figure 2 – Proportion of patients who recognized the flavors**

**4. Conclusion**

We believe that the proposed survival curve approach can serve as an alternative procedure to statistical test for comparison of samples from ordinal qualitative variables. In addition, this new methodology may be more familiar to medical researchers.